\documentclass[12pt]{article}

\usepackage{graphicx}

\newcommand{\sect}[1]{\setcounter{equation}{0}\section{#1}}

\def\be{\begin{equation}}
\def\ee{\end{equation}}
\def\ba{\begin{eqnarray}}
\def\ea{\end{eqnarray}}
\def\nn{\nonumber \\}

\title{Charged Randall-Sundrum black holes and ${\cal N}=4$ super Yang-Mills in $AdS_2 \times S^2$}
\author{Alexander Kaus and Harvey S. Reall \\ \\ Department of Applied Mathematics and Theoretical Physics\\ Centre for Mathematical Sciences \\ Wilberforce Road, Cambridge CB3 0WA, UK \\ ak417, hsr1000@cam.ac.uk}

\begin{document}

\maketitle

\begin{abstract}
We obtain some exact results for black holes in the Randall-Sundrum model with a single brane. We consider an extreme black hole charged with respect to a Maxwell field on the brane. The near-horizon geometry is determined. The induced metric on the brane and the black hole entropy are compared with the predictions of 4d General Relativity. There is good agreement for large black holes, with calculable subleading corrections. As a separate application, the bulk solution provides a gravitational dual for (strongly coupled, large $N$) ${\cal N}=4$ SYM in $AdS_2 \times S^2$ for arbitrary relative size of $AdS_2$ and $S^2$.
\end{abstract}

\sect{Introduction}

In 1999, Randall and Sundrum (RS) discovered a new way of reconciling extra dimensions with observation \cite{RS2}. In their model, our Universe is a 3+1 dimensional brane living in a 4+1 dimensional bulk with a negative cosmological constant. The bulk solution is locally anti-de Sitter space (AdS) and the direction transverse to the brane is non-compact. Nevertheless, at low enough energy, perturbative Newtonian gravity is recovered on the brane at distances large compared to the AdS length $\ell$ \cite{RS2,perturb}.

Subsequently, there was considerable interest in examining whether the agreement with 4d gravity extends beyond perturbation theory. In particular, the question of whether or not the RS model can reproduce the predictions of 3+1 dimensional General Relativity concerning black holes has stimulated a lot of work \cite{gregory}. To answer this question, it is necessary to construct an exact solution in the RS model that describes a black hole localized on the brane. Unfortunately, no satisfactory solution has been obtained analytically. Numerical work \cite{numerical} suggests the existence of such solutions, but only for black holes small compared to $\ell$, for which one does not expect agreement with GR anyway.

The difficulties involved in constructing such a solution can be understood by appealing to the AdS/CFT correspondence \cite{maldacena}. It has been argued \cite{gubser} that AdS/CFT implies that the RS model is equivalent to a 4d effective theory consisting of General Relativity coupled to any matter fields present on the brane {\it and} a conformal field theory (CFT), specifically ${\cal N}=4$ $SU(N)$ super Yang-Mills at large $N$ and strong coupling, with an ultraviolet cut-off. The deviation from 4d GR is therefore attributed to the presence of this CFT: the expectation value of the CFT stress tensor appears as an additional term on the RHS of the effective 4d Einstein equation. A RS black hole solution would be a {\it quantum corrected} black hole solution of the 4d effective theory \cite{efk}. According to this picture, a black hole on the brane would be expected to emit Hawking radiation into the CFT degrees of freedom and therefore would not be static \cite{tanaka,efk}. Hence the difficulty in obtaining, even numerically, a RS black hole solution analagous to the 4d Schwarzschild solution has been attributed to the fact that such a solution would necessarily be time-dependent. However, it has been suggested that strong coupling effects may invalidate this argument \cite{wiseman}.

One way in which progress has been made is to reduce the space-time dimensionality to a 2+1 dimensional brane in a 3+1 dimensional bulk. In beautiful work, Emparan, Horowitz and Myers (EHM) found an exact solution describing a black hole on a 2+1 dimensional analogue of a RS brane \cite{ehm1}. However, vacuum GR in 2+1 dimensions does not admit black hole solutions. From an AdS/CFT perspective, the existence of the EHM solution arises from quantum ``corrections'' due to the dual CFT in 2+1 dimensions, which turn a classical conical singularity into a regular horizon \cite{efk}. In later work, EHM allowed for a negative induced cosmological constant on the brane \cite{ehm2}. In this case, they constructed brane-world black hole solutions that reproduce many properties of black hole solutions of 2+1 dimensional GR with a negative cosmological constant. However, large black holes are not localized on the brane in this model and therefore behave rather differently from what is expected in the original RS model.
 
In this paper, we shall consider the original RS model with a 4+1 dimensional bulk. We shall make progress by considering black holes that are static, spherically symmetric (on the brane), and {\it charged} with respected to a Maxwell field living on the brane. In the extremal limit, such a black hole would have zero temperature and therefore, in the dual 4d picture, would not Hawking radiate. Therefore one would expect a static solution to exist in this case. Finding such a solution is still very difficult since the bulk will depend on two coordinates.\footnote{
Exact solutions for which the induced metric on the brane is an extremal black hole were constructed in Ref. \cite{cvetic}. However, the physical significance of these solutions (which involve non-trivial bulk fields other than the metric) is unclear because they are not localized on the brane and are nakedly singular in the bulk, just like the non-extremal solution of Ref. \cite{chamblin}.} Therefore we make an additional simplification, which is to concentrate on the {\it near-horizon geometry} of the black hole. In 4d GR, a static, spherically symmetric, black hole of charge $Q$ is described by the Reissner-Nordstrom solution which, in the extremal limit, has near-horizon geometry $AdS_2 \times S^2$ where $AdS_2$ and $S^2$ have equal radii $Q$.

Our strategy is to write down the most general near-horizon geometry for the bulk solution consistent with the symmetries. The bulk Einstein equation reduces to ODEs. An important ingredient in solving these ODEs is {\it regularity}: the bulk must be non-singular. The Einstein equation is straightforward to integrate numerically, yielding a 1-parameter family of solutions. We then solve the Israel junction condition describing the gravitational effect of the brane. This relates the single parameter in the bulk uniquely to the charge $Q$ on the brane. We then have a 1-parameter family of solutions labelled by $Q$.

The induced metric on the brane is $AdS_2 \times S^2$, but with unequal radii $L_1$ and $L_2$ for $AdS_2$ and $S^2$ respectively. For large $Q/\ell$, we shall show that
\be
\label{eqn:Lsol}
 L_1^2 = Q^2 - \frac{3\ell^2}{4} + \ldots, \qquad L_2^2 = Q^2 - \frac{ \ell^2}{4} + \ldots
\ee
where the ellipses denote terms subleading in $Q/\ell$. Hence we have agreement with classical GR for large $Q/\ell$. Furthermore, from the bulk near-horizon geometry, we can determine an important property of the full black hole solution, namely the entropy as a function of charge. We find that that 5d Bekenstein-Hawking entropy is (taking $Q>0$ henceforth)
\be
\label{eqn:entropyexp}
 S_5 = \frac{\pi \ell Q^2}{G_5} - \frac{\pi \ell^3}{G_5} \log \left(\frac{Q}{\ell} \right) + \ldots,
\ee
The first term is the usual 4d Bekenstein-Hawking entropy of an extremal RN black hole (since $G_4 = G_5/\ell$ in the RS model). Hence the entropy differs from the 4d result by a logarithmic correction that is subleading for large $Q/\ell$.  We can also determine the proper length of the horizon in the direction transverse to the brane. This is
\be
\label{eqn:rho0soln}
 \rho_0 = \ell \log \left( \frac{Q}{\ell} \right) + \ell \log 2 + \ldots
\ee
Hence we find agreement with the behaviour $\rho_0 \propto \ell \log (L_2/\ell)$ predicted by EHM \cite{ehm1} using an argument based on the instability of horizons with larger $\rho_0$ \cite{chamblin}.

These large $Q$ results are obtained {\it analytically} by solving the Einstein and Israel equations at large ``radius'' in the bulk. The numerical results are required only to confirm that the solutions are globally regular.

Recall that, from the dual 4d perspective, our solution is a solution of the 4d Einstein equations coupled to a Maxwell field and the strongly coupled CFT with UV cut-off $\ell^{-1}$. We can translate the above results into 4d language using $\ell^3/G_5 = 2N^2/\pi$ \cite{maldacena}. This gives
\be
 L_1^2 = Q^2 - \frac{3 G_4 N^2}{2 \pi} + \ldots, \qquad L_2^2 = Q^2 - \frac{G_4 N^2}{2\pi} + \ldots, \qquad S_5 = \frac{\pi Q^2}{G_4} - 2N^2 \log \left(\frac{Q}{\ell} \right) + \ldots
\ee
These results are in agreement qualitatively with previous analyses of quantum corrections to black holes at {\it weak} coupling. In particular, 1-loop corrections to black hole entropy arising from free conformally coupled fields give a logarithmic term \cite{logs}.\footnote{These results do not take account of the gravitational backreaction of the quantum matter whereas ours do. However, this backreaction results in a constant shift in the 4d horizon area (as can be seen from the formula for $L_2^2$) which is subleading compared to the logarithmic term arising from the entropy in the quantum matter fields.} Ref. \cite{solodukhin} has studied the particular case of quantum corrections arising from a free massless scalar field to the classical $AdS_2 \times S^2$ solution. The $Q$-dependence of the corrections is the same as we have found.\footnote{In Ref. \cite{solodukhin}, the correction to $L_2^2$ was ${\cal O}(1)$, as we have found, but there was no ${\cal O}(1)$ correction to $L_1^2$. This is probably just an ``accident'' that occurs for the particular case of a free scalar.} In summary, our results are of the same form as arises from quantum corrections due to ${\cal O}(N^2)$ free conformally coupled fields. There is no indication that strong coupling leads to qualitatively new physics.

We can also consider small black holes, i.e., ones with $Q/\ell \ll 1$. We find that $L_1,L_2 \sim (\ell Q^2)^{1/3}$. The entropy behaves as $S_5 \sim \ell Q^2/G_5$, i.e., the same as for large black holes, but with a smaller coefficient.

Finally, we note that some of our bulk solutions are asymptotically locally AdS, with conformal boundary $AdS_2 \times S^2$. The ratio of the radii of $AdS_2$ and $S^2$ corresponds to the one free parameter in the bulk. We have found the only regular solutions preserving the symmetries of $AdS_2\times S^2$. Hence they must provide the gravitational solutions dual to the vacuum state of ${\cal N}=4$ SYM in $AdS_2 \times S^2$ for arbitrary radii. 

\sect{Near-horizon geometry of charged braneworld black hole}

\subsection{The bulk}

Consider a static brane-world black hole. It is natural to assume that the bulk
will also be static. In the bulk, the surface gravity is constant. Hence, by
continuity, it will take the same value on the brane. Therefore if the horizon
is degenerate on the brane then it will also be degenerate in the bulk. We can
take a near-horizon limit. It was proven in \cite{klr} that the near-horizon
geometry of a static extreme black hole can be written in the warped product
form
\be
 ds^2 = A(x)^2 d\Sigma^2 + g_{ab}(x) dx^a dx^b,
\ee
where $d\Sigma^2$ is the metric on a 2d Lorentzian space $M_2$ of constant
curvature (i.e. Minkowski, or (anti)-de Sitter spacetime)  and $g_{ab}$ is the metric on a spatial cross-section of the horizon.
In general, the warped product structure is only local, but if the horizon is simply connected then it is  global \cite{klr}. In our case, we are interested in a spherical black hole on the brane. The topology of the part of
the black hole horizon lying on either side of the brane will be a hemisphere
of $S^3$, i.e., a 3-ball, which is simply connected. Hence the metric on either side of the brane is globally a warped product.

We now restrict attention to a spherically symmetric black hole, i.e., there is
a $SO(3)$ symmetry with $S^2$ orbits. We can then choose coordinates $x^a =
(\rho,\theta,\phi)$ so that
the near-horizon geometry is
\be
\label{eqn:bulk}
 ds^2 = A(\rho)^2 d\Sigma^2 + d\rho^2 + R(\rho)^2 d\Omega^2,
\ee
where $A$ and $R$ are non-negative functions. By rescaling $A$ we can arrange
for $M_2$ to have Ricci scalar $2k$ with $k \in \{-1,0,1\}$. The bulk Einstein
equation is
\be
 R_{\mu\nu} = -\frac{4}{\ell^2} g_{\mu\nu},
\ee
where $\ell$ is the AdS radius of curvature. The near-horizon metric is
cohomogeneity-1, so the Einstein equation reduces to ODEs. Explicitly, these
are: 
\be
\label{eqn:Aeq} 
\frac{k}{A^2}-\frac{{A'}^2}{A^2}-\frac{2A' R'}{A R}-\frac{A''}{A}=-\frac{4}{\ell^2}
\ee
\be
 \frac{A''}{A}+\frac{R''}{R}=\frac{2}{\ell^2}
\ee
\be
\label{eqn:Req}
 \frac{1}{R^2}-\frac{{R'}^2}{R^2}-\frac{2A'R'}{AR}-\frac{R''}{R}=-\frac{4}{\ell^2}.
\ee
Adding these equations gives the Hamiltonian constraint:
\be
\label{eqn:hamiltonian}
 \frac{k}{A^2} + \frac{1}{R^2} = \frac{{A'}^2}{A^2} + \frac{{R'}^2}{R^2} + \frac{4 A'R'}{AR} - \frac{6}{\ell^2}.
\ee
In the bulk, compactness of the horizon implies that $R(\rho)$ must vanish
somewhere. We can shift $\rho$ so that this occurs at $\rho=0$, with
$R(\rho)>0$ for $\rho>0$. Smoothness at $\rho=0$ requires that $A(\rho)$ has a
Taylor series consisting of even powers of $\rho$ and $R(\rho)$ a Taylor series
consisting of odd powers of $\rho$, with
\be
 A(0) \equiv A_0 > 0, \qquad R'(0)=1.
\ee
If $k=0$ then there is no loss of generality in setting $A_0=1$ so these ``initial'' data are unique. However, if $k=\pm 1$ then we have a 1-parameter family of initial data for the integration of the Einstein equation. Note that $R$ is monotonically increasing in the bulk: if it were not then $R$ would have a local maximum, where $R'=0$, and evaluating (\ref{eqn:Req}) at this point gives $R''>0$, which is impossible at a local maximum.

We have not been able to determine the general solution of the above equations analytically. However, the behaviour for $\rho \ll \ell, A_0$ can be determined by solving using a series expansion:
\ba
\label{eqn:series}
 A &=& A_0 + \left(\frac{k}{A_0} + \frac{4A_0}{\ell^2} \right) \frac{\rho^2}{6} + \left(-\frac{11k^2}{A_0^3} - \frac{40 k}{\ell^2 A_0} + \frac{16 A_0}{\ell^4} \right) \frac{\rho^4}{1080} + \ldots \nn
 R &=& \rho + \left(-\frac{k}{A_0^2} + \frac{2}{\ell^2} \right) \frac{\rho^3}{18} + \left(\frac{53k^2}{A_0^4} + \frac{220 k}{\ell^2 A_0^2} + \frac{212}{\ell^4} \right) \frac{\rho^5}{5400} + \ldots
\ea 
Some solutions of the above form have been discussed previously. If $k=0$ then the solution corresponds to the metric dual to the ground state of ${\cal N}=4$ SYM on $R \times S^1 \times S^2$ obtained in Ref. \cite{copsey}, although it is not written in a  manifestly Poincar\'e invariant form there. For $k=+1$, we can analytically continue $M_2=dS_2$ to $S^2$, giving a Riemannian metric of the form discussed by B\"ohm, who proved that the above method leads to complete metrics on $R^3 \times S^2$ \cite{bohm2}.

\subsection{The brane}

We take the brane to have action
\be
 S_{\rm brane} = \int d^4 z \sqrt{-h} \left( -\sigma - \frac{1}{16\pi G_4}
F_{ij}F^{ij} \right),
\ee
where $\sigma$ is the brane tension, $G_4$ is the Newton constant on the brane,
and $F$ is the Maxwell field on the brane. We shall set the brane
tension to the Randall-Sundrum value $\sigma=3/(4\pi G_5\ell)$, which gives
$G_4 = G_5/\ell$. The Israel junction condition for a $Z_2$-symmetric brane is
\be
 K_{ij} = \frac{1}{\ell} h_{ij} + \ell \left(F_i{}^k F_{jk} - \frac{1}{4}
h_{ij} F^{kl} F_{kl} \right),
\ee
where $K_{ij}$ is the extrinsic curvature of the brane and $h_{ij}$ its induced
metric.
We assume the brane is located at $\rho=\rho_0$, so the induced metric on the
brane is a product $M_2 \times S^2$:
\be
\label{eqn:4dmetric}
 ds_4^2 = L_1^2 d\Sigma^2 + L_2^2 d\Omega^2,
\ee
where
\be
 L_1 \equiv A(\rho_0), \qquad L_2 \equiv R(\rho_0)
\ee
are the radii of $M_2$ and $S^2$ respectively.
We assume the Maxwell field to be spherically symmetric. By a duality rotation
we may take it to be purely electric, i.e.,
\be
 \star_4 F = Q d\Omega,
\ee
where $Q$ is the total electric charge:
\be
 Q = \frac{1}{4\pi} \int_{S^2} \star F.
\ee
Note that this will agree with the electric charge defined by a surface
integral at infinity on the brane in the full (asymptotically flat) black hole
spacetime. We shall take $Q \ge 0$ henceforth.

The Israel junction condition implies that we must keep the region $0 \le \rho \le \rho_0$ of the bulk, and satisfy the boundary conditions
\be
\label{eqn:israel}
\frac{A'(\rho_0)}{A(\rho_0)} = \frac{1}{\ell} - \frac{\ell Q^2}{2 R(\rho_0)^4}, \qquad
 \frac{R'(\rho_0)}{R(\rho_0)} = \frac{1}{\ell} + \frac{\ell Q^2}{2 R(\rho_0)^4}.
\ee
We can use these to evaluate the Hamiltonian constraint (\ref{eqn:hamiltonian}) at $\rho=\rho_0$, giving
\be
 \frac{k}{L_1^2} + \frac{1}{L_2^2} = - \frac{\ell^2 Q^4}{2 L_2^8},
\ee
from which we deduce $k=-1$, so the metric on the brane is $AdS_2 \times S^2$, in agreement with 4d GR. However, this equation also implies $L_1 < L_2$, so the $S^2$ radius is greater than the $AdS_2$ radius, in contrast to 4d GR (which predicts $L_1=L_2=Q$). Note that the geometric interpretation of $\rho_0$ is as the proper length of the horizon in the direction transverse to the brane.

\subsection{The solutions}

We take $k=-1$ henceforth.
Our strategy will be to fix $A_0$ and integrate the bulk Einstein equations to
determine $A(\rho)$ and $R(\rho)$. We do this by using the series expansions (\ref{eqn:series}) to fix initial data at $\rho=\epsilon \ll A_0, \ell$ (we can't start at $\rho=0$ because the equations are singular there) and then evolve the solution using the second order equations (\ref{eqn:Aeq}) and (\ref{eqn:Req}). The Hamiltonian constraint (\ref{eqn:hamiltonian}) is used to monitor the accuracy of the solution. We then choose the two parameters $\rho_0$ and $Q$ so that the two components of the Israel junction condition are satisfied. This will give a 1-parameter family of solutions.

We start by noting the existence of two special bulk solutions that can be determined analytically. First, if $A_0 = \ell/2$ we find
\be
 A \equiv \frac{\ell}{2}, \qquad R = \frac{\ell}{\sqrt{2}} \sinh \left( \frac{\sqrt{2} \rho}{\ell} \right).
\ee 
The bulk metric is then $AdS_2 \times H^3$. The other special case has $A_0 = \ell$, giving
\be
 A = \ell \cosh \left( \frac{\rho}{\ell} \right), \qquad R = \ell \sinh \left( \frac{\rho}{\ell} \right),
\ee
which is just $AdS_5$ written in coordinates adapted to a foliation by $AdS_2 \times S^2$ hypersurfaces. For the $AdS_5$ solution, the Israel equation cannot be satisfied.
For the $AdS_2 \times H^3$ solution we find that the Israel equations are satisfied if
\be
 \sinh\left( \frac{\sqrt{2} \rho_0}{\ell} \right) = 1, \qquad Q = \frac{\ell}{\sqrt{2}}.
\ee
We then have
\be
 L_1 = \frac{\ell}{2}=\frac{Q}{\sqrt{2}}, \qquad L_2 = \frac{\ell}{\sqrt{2}}=Q.
\ee
Hence, for this (small) black hole, the near-horizon metric on the brane differs from the corresponding solution in 4d GR only through the fact that the radius of the $AdS_2$ is $Q/\sqrt{2}$ rather than $Q$.

\begin{figure}
\begin{center}
 \includegraphics[bb=0 0 240 163]{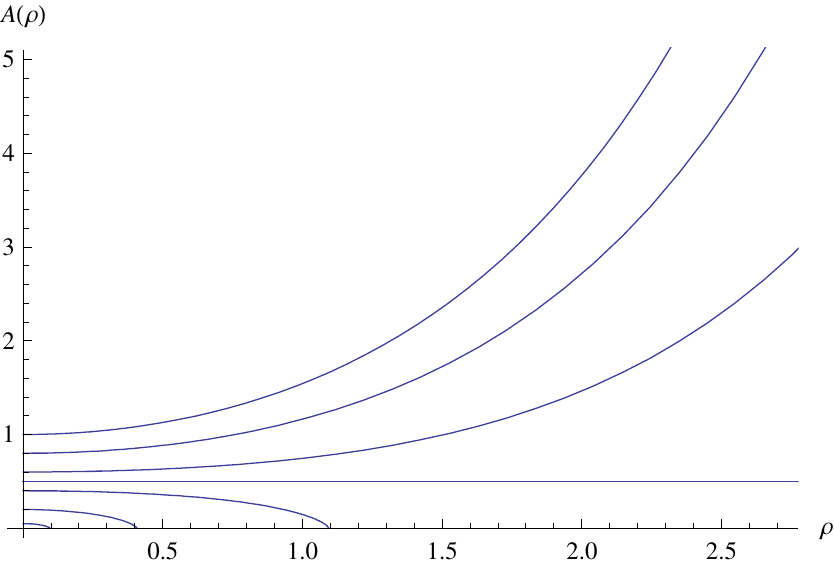}
 \includegraphics[bb=0 0 240 163]{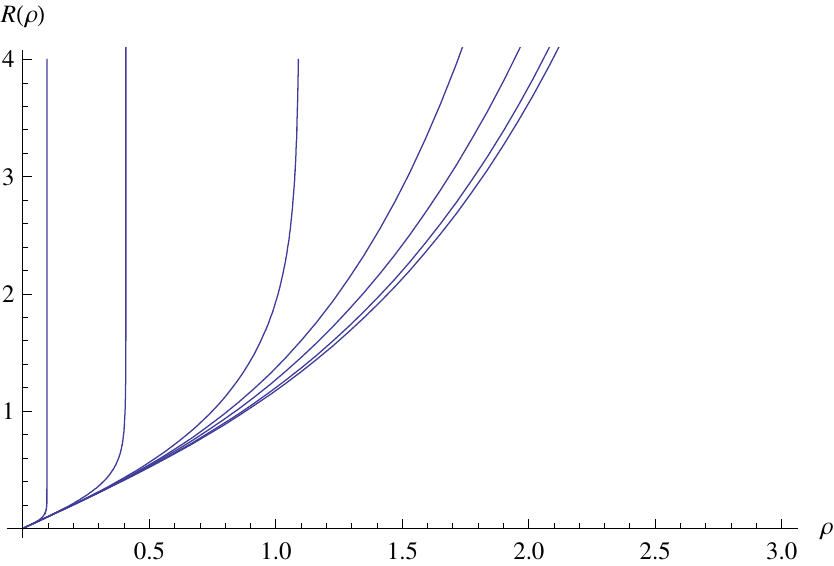}
\end{center}
\caption{Bulk solutions for $A_0 = 0.05,0.2,0.4,0.5,0.6,0.8,1.0$ (from bottom to top on left plot, from left to right on right plot, units $\ell=1$).}
\label{fig:bulksoln}
\end{figure}

Now we shall discuss more general values for $A_0$. First we shall describe the qualitative beheaviour of the bulk solution. The behaviour of $A$ and $R$ for different values of $A_0$ is shown in Fig. \ref{fig:bulksoln}.
We argued above that $R$ must increase monotonically. For $0<A_0 < \ell/2$, we find that $A$ decreases montonically, and vanishes at some finite value $\rho=\rho_1$. $R$ diverges at $\rho=\rho_1$. A calculation of the square of the Riemann tensor reveals that $\rho=\rho_1$ is a curvature singularity.

For $A_0 > \ell/2$, we find that both $A$ and $R$ increase monotonically, and are proportional to $\exp(\rho/\ell)$ for large $\rho$, indicating that the bulk solution is asymptotically locally $AdS_5$ as $\rho \rightarrow \infty$, with conformal boundary $AdS_2 \times S^2$. The solution is topologically trivial. 
On the conformal boundary, the ratio $a$ of the radius of the $AdS_2$ to that of the $S^2$ is a monotonically increasing function of $A_0$, with $a \rightarrow 0$ as $A_0 \rightarrow \ell/2$, $a \rightarrow \infty$ as $A_0 \rightarrow \infty$, and $a=1$ for the $AdS_5$ bulk ($A_0 = \ell$). Our bulk solution must provide the gravitational dual of the ground state of ${\cal N}=4$ SYM in $AdS_2 \times S^2$ for arbitrary $a$ (since it is the only regular solution with the appropriate symmetries).

The Israel junction condition can be satisfied if, and only if, $0<A_0 < \ell$, when it determines $\rho_0$ and $Q$ in terms of $A_0$. For $0<A_0<\ell/2$, we find $\rho_0<\rho_1$, so the curvature singularity at $\rho=\rho_1$ is not present in the spacetime containing the brane. We find that $\rho_0$ and $Q$ are monotonically increasing functions of $A_0$ which vanish as $A_0 \rightarrow 0$ and diverge as $A_0 \rightarrow \ell$. Physically, it is more interesting to take $Q$, rather than $A_0$ as the dependent variable, and we shall do so henceforth.

\begin{figure}
\begin{center}
 \includegraphics[bb=0 0 240 152]{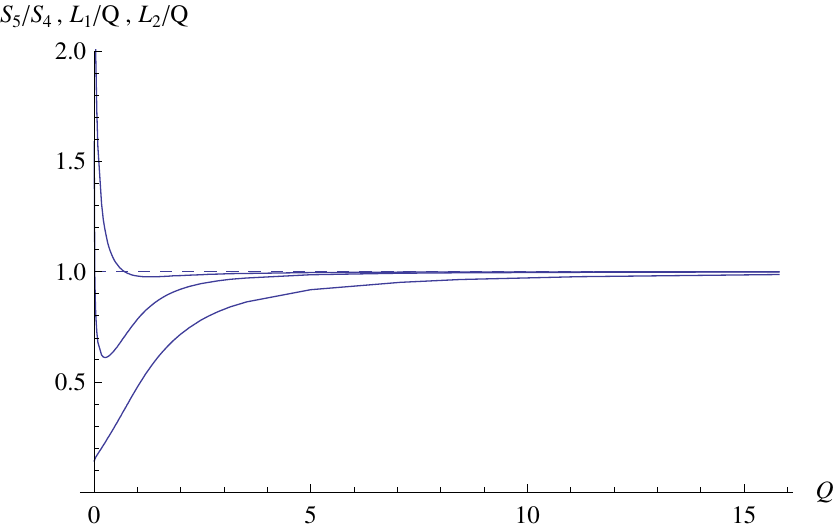}
 \includegraphics[bb=0 0 240 162]{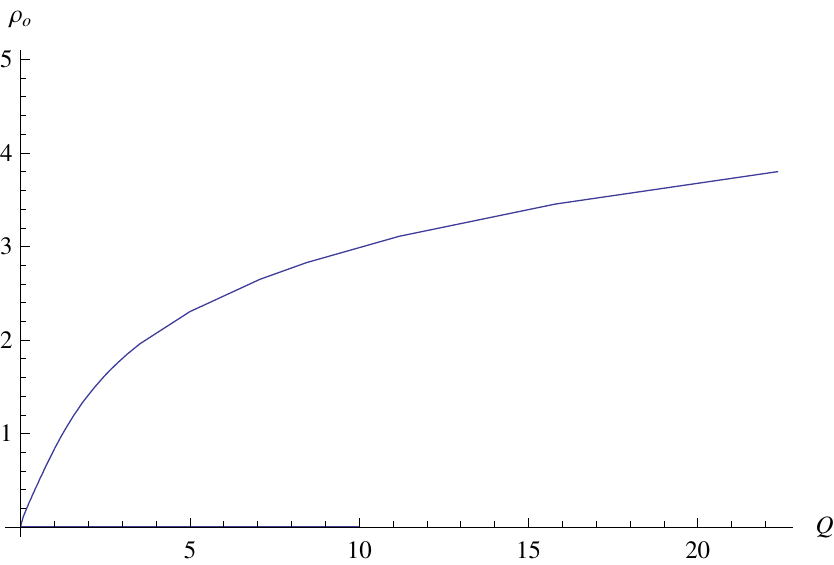}
\end{center}
\caption{Left: $L_2/Q$ (top), $L_1/Q$ (middle) and $S_5/S_4$ (bottom). Note that the first two curves diverge at small $Q$. Right: $\rho_0$, the proper length of the horizon transverse to the brane. (Units $\ell=1$.)}
\label{fig:L1L2rho0}
\end{figure}

Figure \ref{fig:L1L2rho0} shows how $L_1$, $L_2$ and $\rho_0$ depend on $Q$.
$L_1/Q$ and $L_2/Q$ both approach $1$ for large $Q/\ell$. Hence the induced geometry on the brane agrees with the prediction of 4d GR for large black holes. $\rho_0$ grows as $\ell \log (Q/\ell) \approx \ell \log (L_2/\ell)$ for large $Q$, in agreement with general expectations of brane-world black holes \cite{ehm1}.

The Bekenstein-Hawking entropy of the solution is determined from the area of the event horizon:\footnote{Note that taking account of the bulk on both sides of the brane contributes an overall factor of $2$.}
\be
\label{eqn:bulkentropy}
 S = \frac{2 \pi}{G_5} \int_0^{\rho_0} R(\rho)^2 d\rho.
\ee
The usual 4d Bekenstein-Hawking entropy of an extremal RN black hole of charge $Q$ is
\be
 S_4 = \frac{\pi Q^2}{G_4} = \frac{\pi \ell Q^2}{G_5}.
\ee
For a large black hole, the integral in (\ref{eqn:bulkentropy}) is dominated by the contribution from $\rho \approx \rho_0$ (as anticipated in \cite{ehm1}), where $R(\rho) \approx L_2 \exp((\rho-\rho_0)/\ell)$. We then find $S \approx S_4$ upon using $L_2 \approx Q$. The ratio $S_5/S_4$ is shown in figure \ref{fig:L1L2rho0}. It tends to $1$ at large $Q/\ell$, demonstrating agreement with 4d GR.

The behaviour for small and large black holes can be understood analytically, as we show in the next two subsections.

\subsection{Small black holes}

For a black hole much smaller than the $AdS$ scale $\ell$, we can neglect 
the cosmological constant in Einstein's equation. The only dimensionful 
parameter in the bulk is $A_0$ so dimensional analysis gives $A = A_0 
\hat{A}(\rho/A_0)$ and $R=\rho \hat{R}(\rho/A_0)$ for some dimensionless 
functions $\hat{A},\hat{R}$. The sum of equations (\ref{eqn:israel}) 
implies that $\rho_0 \sim A_0$. Hence all lengths in the problem scale as 
$A_0$. The difference of equations (\ref{eqn:israel}) now implies that 
$A_0 \sim (\ell Q^2)^{1/3}$. Hence $L_1,L_2,\rho_0 \sim (\ell Q^2)^{1/3}$, 
i.e., the black hole behaves like a 5d black hole of radius $(\ell 
Q^2)^{1/3}$. Our numerical results are $L_1 \approx 0.24 (\ell 
Q^2)^{1/3}$, $L_2 \approx 0.66 (\ell Q^2)^{1/3}$, $\rho_0 \approx 0.56 
(\ell Q^2)^{1/3}$. We also have $S_5 \sim (\ell Q^2)/G_5 \propto S_4$, 
just as for a large black hole, however the coefficient is now smaller: 
$S_5/S_4 \approx 0.14$.

\subsection{Large black holes}

Large black holes have $A_0 \approx \ell$ so if the brane were not present then the bulk solution would be asymptotically locally AdS, and the metric would be close to the $AdS_5$ metric (which arises for $A_0 = \ell$). We can calculate analytically many properties of these black holes by solving the Einstein equation near the $AdS_2 \times S^2$ conformal boundary of the spacetime. The latter is specified by the ratio $a$ of the radius of $AdS_2$ to that of $S^2$. The asymptotic solution can be determined using the results of Ref. \cite{skenderis}. Setting $\ell=1$, we find that
\be
 ds^2 =\frac{dr^2}{4r^2} + A(r)^2 d\Sigma^2 + R(r)^2 d\Omega^2,
\ee
where
\be
\label{eqn:Aexp}
A(r) = a^2 \left[ \frac{1}{r} + \frac{1}{6} + \frac{1}{3a^2} - \frac{1}{48} \left( 1 - \frac{1}{a^4} \right) r \log r + \left( \frac{5}{288} + \frac{1}{36 a^2} + \frac{5}{288 a^4} + \lambda \right) r + \ldots \right]
\ee
\be
R(r) =  \frac{1}{r} - \frac{1}{3} - \frac{1}{6a^2} + \frac{1}{48} \left( 1 - \frac{1}{a^4} \right) r \log r + \left( \frac{5}{288} + \frac{1}{36 a^2} + \frac{5}{288 a^4} - \lambda \right) r + \ldots
\ee
The coordinate $r$ is related to our previous coordinate $\rho$ by 
\be
r = r_0 e^{-2\rho}
\ee
for some constant $r_0$. The conformal boundary is at $r=0$. The asymptotic solution involves an unknown constant $\lambda$ which is to be determined by the requirement of bulk regularity. The ellipses denotes terms that are subleading in $r$ relative to the terms written above. These are uniquely determined by $a,\lambda$.

Now assume that the brane is at $r=\epsilon$. The extrinsic curvature is
\be
 K_{ij}dx^i dx^j = K_1 d\Sigma^2 + K_2 d\Omega^2,
\ee
where
\ba
K_1 &=&
a^2 \left\{ \frac{1}{\epsilon} + \frac{1}{48} \left( 1 - \frac{1}{a^4} \right) \epsilon \log \epsilon + \left[ \frac{1}{48} \left( 1- \frac{1}{a^4} \right) - \frac{5}{288} - \frac{1}{36}{a^2} - \frac{5}{288 a^4} - \lambda  \right] \epsilon + \ldots \right\}  \nn
K_2 &=&  \frac{1}{\epsilon} - \frac{1}{48} \left( 1 - \frac{1}{a^4} \right) \epsilon \log \epsilon + \left[- \frac{1}{48} \left( 1- \frac{1}{a^4} \right) - \frac{5}{288} - \frac{1}{36}{a^2} - \frac{5}{288 a^4} + \lambda  \right] \epsilon + \ldots
\ea
The Israel equation gives
\be
\label{eqn:K1K2}
 \frac{K_1}{L_1^2} = 1 - \frac{Q^2}{2L_2^4}, \qquad \frac{K_2}{L_2^2} =  1 + \frac{Q^2}{2L_2^4},
\ee
where $L_1 \equiv A(\epsilon)$, $L_2 \equiv R(\epsilon)$. Expanding the sum of these equations as a series in $\epsilon$ gives
\be
 a^2 = 1 - \frac{3}{2} \epsilon + \ldots
\ee
We then have
\be
 L_1^2 = \frac{1}{\epsilon} - 1 + \ldots, \qquad L_2^2 = \frac{1}{\epsilon} - \frac{1}{2} + \ldots
\ee
The difference of equations (\ref{eqn:K1K2}) gives
\be
 Q^2 = \frac{1}{\epsilon} + 4\lambda - \frac{1}{4} + \ldots
\ee
The constants $\lambda$ and $r_0$ must be fixed by bulk regularity. We explained above that there is a 1-parameter family of regular bulk solutions hence $\lambda=\lambda(a)$, $r_0 = r_0(a)$. Note that $a \rightarrow 1$ as $\epsilon \rightarrow 0$. The unique regular bulk solution with $a=1$ is the $AdS_5$ solution. A calculation reveals that this solution has $\lambda = 0$ and $r_0=4$. This fixes the leading term in the $\epsilon$ expansions of $\lambda$ and $r_0$:
\be
 \lambda |_{\epsilon=0}=0, \qquad r_0 |_{\epsilon=0} = 4.
\ee
Hence
\be
\label{eqn:Qeps}
 Q^2 = \frac{1}{\epsilon} - \frac{1}{4} + \ldots
\ee
Inverting this determines $\epsilon$ as a function of $Q$. Plugging this into the expressions for $L_1$ and $L_2$ then gives the solutions (\ref{eqn:Lsol}) presented in the introduction. We also have
\be
 \rho_0 = \frac{1}{2} \log \left( \frac{r_0}{\epsilon} \right) = \frac{1}{2} \log \left( \frac{1}{\epsilon} \right) + \log 2 + \ldots
\ee
Eliminating $\epsilon$ using (\ref{eqn:Qeps}) then gives the solution (\ref{eqn:rho0soln}).

The Bekenstein-Hawking entropy is
\be
 S_5 = \frac{\pi}{G_5} \int_{\epsilon}^{r_0} \frac{R(r)^2}{r}  dr,
\ee
Differentiating with respect to $\epsilon$, then using the above series solution for $L_2 = R(\epsilon)$ gives
\be
 \frac{dS_5}{d\epsilon} = \frac{\pi}{G_5} \left( - \frac{1}{\epsilon^2} + \frac{1}{2\epsilon} + \ldots \right),
\ee
Integrating, then writing $\epsilon$ in terms of $Q$ gives (\ref{eqn:entropyexp}).

It would be nice to characterize the form of the next-to-next-to leading 
order (NNLO) corrections to our results. To do so would require 
determining the dependence of $\lambda$ and $r_0$ on $\epsilon$, which we 
are unable to do analytically. However, our numerical results suggest 
strongly that the NNLO terms in $L_1^2$, $L_2^2$ are\footnote{ This 
differs from the ${\cal O}(1/Q^2)$ reported in Ref. \cite{solodukhin} for 
quantum corrections to $AdS_2 \times S^2$ arising from a free scalar. This 
may be because our effective 4d theory involves an {\it interacting} CFT.} 
${\cal O}(\log Q/Q^2)$, in $\rho_0$ the NNLO term is ${\cal O}(1/Q^2)$ and 
in $S_5$ it is a constant $\approx -1.4 \ell^3/G_5$.

In summary, we can calculate many properties of large black holes analytically. The numerical and analytical results are in excellent agreement. This analytical approach alone would not be a satisfactory derivation of the behaviour of large black holes because it does not demonstrate that a regular bulk solution actually exists. Demonstrating regularity requires the numerical work above. However, having verified that a regular solution does indeed exist, this analytical approach provides an explanation of its properties.

\section{Discussion}

Our results provide the first quantitative demonstration that the single brane RS model can reproduce the predictions of 3+1 dimensional GR for black hole physics, for black holes large compared to the AdS scale.

We have considered a black hole charged with respect to a Maxwell field on the brane. However, note that our bulk solution is independent of what kind of matter field is present on the brane. Hence the near-horizon geometry of {\it any} static, spherically symmetric, extremal black hole solution should have the same solution in the bulk (although possibly with $k=0$ or $1$). It might be interesting to investigate other extremal black holes on the brane e.g. with additional fields, or an induced cosmological constant on the brane. For example, one could investigate the brane-world analogue of extremal Reissner-Nordstrom-AdS or the Nariai solution ($dS_2 \times S^2$, which would need $k=1$ so the bulk geometry would be one of the solutions of Ref. \cite{bohm2} analytically continued to Lorentzian signature).

As noted in the introduction, the only case in which analytic solutions describing brane-world black holes are available is for a 3+1 dimensional bulk. The same is true for near-horizon geometries of static extremal black holes. This is because the general static near-horizon geometry solution of the Einstein equation (with cosmological constant) in 3+1 dimensions is known \cite{crt}. It is an analytically continued version of the Schwarzschild-(A)dS metric. This metric could be used to construct the near-horizon geometry of static extremal brane-world black holes on a 2+1 dimensional brane.

Note that we could not determine our bulk solution analytically (except in a special case) and had to resort to numerical integration. This suggests that the harder problem of finding the full bulk geometry of an extreme brane-world black hole also will not be possible analytically. It would be interesting to see if this problem could be solved numerically using the methods of Ref. \cite{numerical}.

We have considered only static black holes. Constructing the near-horizon geometry of an extreme rotating black hole on the brane (analagous to extreme Kerr) would be interesting. However, the (near-horizon) bulk solution in this case would be cohomogeneity-2 so finding it is probably as hard as finding a full solution for an extreme static, spherically symmetric, brane-world black hole solution.

Our bulk solutions, multipled by $S^5$, provide the geometries dual to the ground state of (strongly coupled, large $N$) ${\cal N}=4$ SYM  in $AdS_2 \times S^2$ for arbitrary radii (or some other CFT if $S^5$ is replaced by another positive Einstein space $X$). The CFT stress tensor can be calculated using the method of Ref. \cite{skenderis}. (However, the result is renormalization scheme dependent - it involves an arbitrary constant.) We expect that similar geometries could be constructed that are dual to CFTs in $M_2 \times S^3$, $M_3 \times S^2$, $M_3 \times S^3$, $M_2 \times S^4$, and $M_2 \times S^2 \times S^2$, where $M_p$ is a $p$-dimensional Lorentzian space of constant curvature.

\bigskip

\begin{center}{\bf Acknowledgments}\end{center}

\medskip

\noindent We are very grateful R. Emparan for his comments on a draft of this paper. We have enjoyed discussions with G. Gibbons, S. Hawking, R. Monteiro, S. Ross and J. Santos. HSR is a Royal Society University Research Fellow.

\end{document}